# Detecting Gamification in Breast Cancer Apps: an automatic methodology for screening purposes

Guido Giunti
Salumedia Tecnologías
Seville, Spain
guidogiunti@salumedia.com

Diego H. Giunta
Hospital Italiano de Bs. As.
Buenos Aires, Argentina
diego.giunta@hiba.org.ar

Santiago Hors-Fraile
Universidad de Sevilla
Seville, Spain
santiago@atc.us.es

Minna Isomursu
Oulun Yliopisto
Oulu, Finland
minna.isomursu@oulu.fi

Diana Karosevičiūtė
Kauno Tech. Universitetas
Kaunas, Lithuania
diana.karoseviciute@ktu.edu

*Abstract*—Breast cancer is the most common cancer in women both in developed and developing countries. More than half of all cancer mobile application concern breast cancer. Gamification is widely used in mobile software applications created for health-related services. Current prevalence of gamification in breast cancer apps is unknown and detection must be manually performed. The purpose of this study is to describe and produce a tool allowing automatic detection of apps which contain gamification elements and thus empowering researchers to study gamification using large data samples. Predictive logistic regression model was designed on data extracted from breast cancer apps' title and description text available in app stores. Model was validated comparing estimated and benchmark values, observed by gamification specialists. Study's outcome can be applied as a screening tool to efficiently identify gamification presence in breast cancer apps for further research.

*gamification; mHealth; health apps; medical apps; medical informatics; breast cancer ; predictive models*

## I. INTRODUCTION

Breast cancer is the most common cancer in women both in developed and developing countries. It is estimated that worldwide over half a million women died in 2011 due to breast cancer [1]. Fortunately, thanks to advancements in treatments, breast cancer survivorship is on a steady rise [2], [3]. This cancer is no longer thought of as an acute illness but rather a chronic condition with focus on long term goals and wellbeing promotion [4]. Survivors of breast cancer represent a unique group who must be aware of the long-term consequences of their treatment and be given information to encourage a proactive approach to their overall health [4], [5]. Breast cancer survivors require a personalized needs assessment; a self-management based care approach as well as individualized follow-up and support [6].

The use of mobile software applications (apps) for health and wellbeing promotion has grown exponentially in recent years [7]. Mobile health (mHealth) is defined as the delivery of healthcare or health related services through the use of portable devices [8]. There are currently thousands of healthcare related mobile software applications (apps) available through app stores [9]. Bender et al [10] explored the distribution of cancer apps across the four major smartphone platforms in 2013; this study found that half of the mHealth apps (45%) were about breast cancer. The use of gamification in mHealth apps is now a popular strategy [11]–[16] but no data is available on the prevalence of gamification in breast cancer apps. Gamification is often defined as "the use of game design elements in non-game contexts" [17]. This definition seems to cover anything from the inclusion of elements frequently associated with games like points, badges and leaderboards to game design thinking applied to reworking and improving work processes. The broadness of the term makes the study of gamification in health apps difficult. Researchers must manually review each application thoroughly to detect the presence of gamification. This approach is operator-dependent and cumbersome for large samples.

There is no current way to identify a priori health apps that contain gamification elements and this task remains a manual labor [16]. Tailoring gamification to the task and intended audience is key to successful gamification [18]. Researchers who want to compare health apps to assess whether gamification leads to a more successful app adoption or track gamification use in Breast Cancer apps have to rely on developers openly declaring that their app uses gamification. Using gamification in Breast Cancer applications presents a challenge because of the cultural connotations that a cancer diagnosis carries which will likely influence the design process and how explicit developers are with the use of gamification. The use of gamification as a distinguishing factor or selling point is not common practice in health applications: a health app for kids with gamification will likely be advertised as "gamified" but one aimed for adults may not.

Our study describes the steps taken to create and validate a predictive model to automatically detect gamification in large samples of breast cancer apps using only an app's title and description text available in app stores. We empirically studied the presence of gamification in breast cancer apps and developed a method that allows for further exploration. The intention of this method is to provide a screening tool for researchers and designers so that further analysis of gamification in breast cancer can be undertaken.

## II. METHODS

### A. Study Population

A cross-sectional of health apps was obtained from the iTunes App Store [19] and Google Play [20] store of the United States of America on February 24th 2016. The study flow is illustrated in Figure 1. The iTunes App Store serves as the official app store for iOS and has 2 million apps available as of June 2016 [21]. Google Play store (originally the Android Market) serves as the official app store for the Android operating system with over 2.2 million apps available as of June 2016 [21]. According to a report, there are more than 165,000 mobile health apps in total on both markets [22].

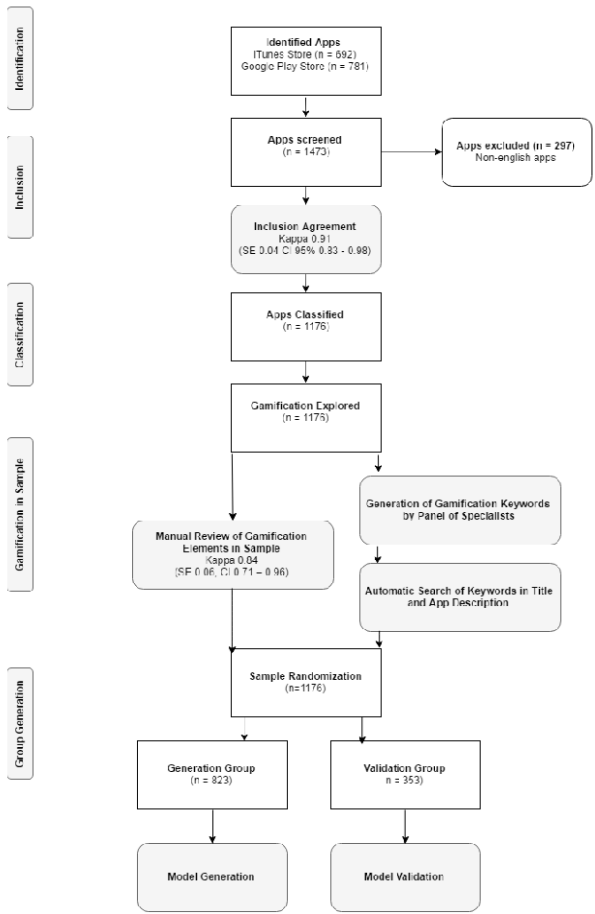

Figure 1. Study Flow.

### B. Data Set Generation

Both stores were systematically searched for "breast cancer" apps on February 24th 2016 using the audience targeting platform 42matters [23] which aggregates mobile applications data and meta-data across the iOS and Google Play stores. All apps that were a partial or complete match for the search terms were initially included. Mobile device fragmentation is a phenomenon that occurs when mobile users from different OS or versions of the same OS need to run running different versions due to different version capabilities or app submission processes to the stores. This caused us to consider basic and "premium" versions of the same app were considered as separate entries as were versions of the same app for different operating systems. This distinction is common practice in systematic app reviews [10].

A small sample (10%) was randomly selected and two reviewers with ample domain experience independently followed the selection criteria discarding non-English apps and sorting them into: a) breast cancer apps: apps specific to breast cancer or breast cancer related conditions (eg Breast Cancer: Beyond the Shock); b) health apps: apps not specific to breast cancer but are health-related (eg My Cancer Coach); and c) miscellaneous apps: apps that may have thematic connections to breast cancer, cancer or health but are not primarily about health (eg Cancer Cause support skins for music players).

In order to assess clarity of the selection criteria, inter-rater reliability was determined using Fleiss Cohen's Inter-rater Reliability Coefficient (kappa) according to Landis & Koch's standards [24]. Disagreements were resolved by consensus involving a third reviewer when necessary.

### C. Gamification Keyword List Generation

A panel of gamification specialists and experienced gamification designers was asked to create a list of "gamification keywords" that they would expect to find in descriptions of app that contain gamification elements. Due to the novelty of gamification as a discipline, specialists were defined as individuals with more than 3 years of first hand work experience designing gamification interventions. The panel was composed of an interdisciplinary group of 7 individuals with varying backgrounds: business professionals (2), healthcare professionals (2), information technology professionals (2), and education sciences professionals (1).

### D. Gamification in data set

Two panel members examined all of the included apps' description and title independently. Text descriptions and title were used as this is the information available to users before installing. The two panel members flagged those they suspected might contain gamification. Fleiss' Kappa was calculated and used as our "gold standard" for current gamification in our data set.

Additionally, an automatic search was performed on the apps' description and title looking for the presence=1 or absence=0 of each of the "gamification keywords" and a corresponding variable for each keyword was set accordingly.

### E. Model Generation

The unit of analysis was each application. Our outcome was gamification presence, which was categorized as present or absent, according to whether the panel members concluded the app might contain gamification elements or not.

The sample was randomized to allocate a generation group (two-thirds of the sample) and a validation group (one third of the sample). Using the generation group, univariate association was assessed between each "gamification keyword" and gamification presence, using a logistic regression model. Odds ratios (ORs) were estimated for the presence of gamification with 95% confidence intervals (95% CI).

Multivariate logistic regression was used for generating a predictive model of gamification presence. Significant variables in univariate analysis were incorporated into the model, along with those considered relevant, even though not significant in univariate analysis.

Diagnostic performance of the model was evaluated using receiver operating characteristic (ROC) curves for the predicted values of gamification presence, which were calculated using the estimated predictive model. We used the accuracy approximated by the area under the ROC curve as a goodness-of-fit measure. Different models were compared using Akaike's Information Criterion (AIC). The model with the largest area under the ROC curve and the lowest AIC was selected.

*F. Model Validation*

The selected predictive model was validated, using the validation group, by comparing predicted values with those observed. We stratified every app case from the validation group according to the probability predicted by the model, in groups: first, second, third and fourth quartiles. We calculated the average probability of gamification presence predicted by the model (the average of individual estimated probabilities), and the observed probability (proportion of gamification presence), within each stratum. The predicted and observed probabilities are shown, for each stratum, in Table IV.

*G. Statistical methods*

Categorical variables are presented as absolute and relative frequencies. Continuous variables are presented as mean and standard deviation or median with interquartile range depending on distribution. A p-value of less than 5% was considered statistically significant. Statistical analysis was performed using STATA v13.

## III. RESULTS

*A. Data Set Generation*

A total of 1473 apps matched the search terms of "breast cancer" in both stores, of which 692 matches were from the iTunes App Store and 781 from the Google Play Store. A random sample (n=146) was independently reviewed by two reviewers. Inter-rater reliability kappa was more than acceptable at 0.91 (SE 0.04, CI 95% 0.83 - 0.98). After removing non-English apps the final data set was constructed (n=1176). Table I shows basic characteristics of the resulting data set.

TABLE I. DATA SET BASIC CHARACTERISTICS

|  | Frequency |
|---|---|
| **Operating System** | |
| - Android | 625 (53.15%) |
| - iOS | 551 (46.85%) |
| **App Type** | |
| - Breast Cancer | 599 (50.93%) |
| - Health | 457 (38.86%) |
| - Misc. | 120 (10.21%) |
| **Gamification Presence ╪** | 241 (20.49%) |

╪ according to our specialists' assessment.

*B. Gamification Keyword List Generation*

An initial list of "gamification keywords" deemed likely to be present when describing an application with gamification elements was constructed by the panel. See Table II.

*C. Gamification in data set*

After reviewing our app data set, panel members determined that 20.49% (n=241) might contain gamification elements. The inter-rater reliability between specialists on this was 0.84 (SE 0.06, CI 0.71 – 0.96).

TABLE II. INITIAL GAMIFICATION KEYWORD LIST

| | | | |
|---|---|---|---|
| achievements | fun | measure | reward |
| badges | funny | medals | routine |
| behavior | game | monitor | rules |
| behavioral | games | motivate | score |
| behaviour | gamification | motivation | share |
| challenge | gamified | motivational | skills |
| change | gamify gaming | multiplayer | social |
| collaborative | goal | narrative | socialize |
| competition | goals | network | statistics |
| competitive | habits | planning | stories |
| connect | incentivate | play | story |
| contest | incentive | player | target |
| diary | incentives | points | team |
| engage | leaderboard | progress | teams |
| engagement | level | purpose | track |
| engaging | leveling | quantify | tracking |
| entertaining | log | quest | trivia |
| entertainment | logging | quests | tutorial |
| experience | map | quiz | |
| explore | master | ranking | |

TABLE III. FINAL GAMIFICATION VARIABLE LIST

| **Activity Tracking** | **Entertainment** | **Game Labels** |
|---|---|---|
| *log; logging; measure; monitor; track; tracking* | *entertainment; fun; funny; play; entertaining* | *gamification; gamified; game; games; gamify; gaming* |
| **Engagement** | **Quizzes** | **Player Aspects** |
| *engage; engagement; engaging* | *quiz; trivia* | *multiplayer; player; team; teams* |
| **Diary** | **Change** | **Story** |
| *diary* | *change* | *story* |
| **Progress** | **Purpose** | **Quest** |
| *progress* | *purpose* | *quest* |
| **Routine** | **Statistics** | |
| *routine* | *statistics* | |

Variable names in bold. Keywords encompassed in italics.

*D. Model Generation*

Univariate analysis was applied to the initial gamification keyword list to find statistically significant associations with the app's text description and title. Each keyword was used as a variable for the analysis. Keywords that represented similar concepts were grouped together as new variables and tested with univariate analysis. The resulting variable list with their depending keywords is shown in Table III.

The predictive model was trained using a generation group which included 823 apps with a gamification frequency of 20.66% (n=170). Table V shows the list of included variables with their estimated coefficients. The area under the ROC curve of the predictive model was 0.89 (95% CI: 0.85-0.91) for the generation group (Figure 2).

Using the predictive model in the validation group (n=353), the area under the ROC curve was estimated at 0.85 (95% CI: 0.79–0.91). For the validation group, predicted and observed probabilities were very similar, for each quartile. First and second quartiles were grouped together because of app distribution. The observed and predicted probabilities for each quartile are shown in Table IV.

TABLE IV. OBSERVED AND PREDICTED PROBABILITIES FOR EACH QUARTILE

| Quartile | Number of Observations | Observed Gamification | Predicted Gamification |
|---|---|---|---|
| Q1-Q2 | 204 | 0.044 | 0.048 |
| Q3 | 62 | 0.145 | 0.096 |
| Q4 | 87 | 0.609 | 0.620 |

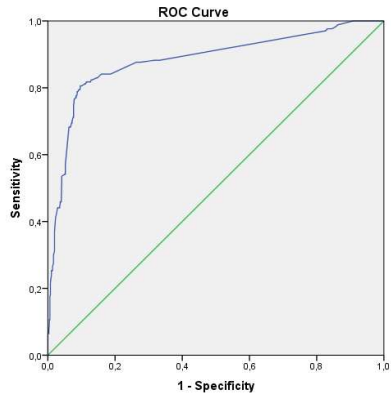

Figure 2. ROC Curve for predictive model in generation group.

TABLE V. VARIABLES OF THE PREDICTIVE MODEL

| Variable | Estimated Coefficient | Standard Error | p | [95% Conf. Interval] | OR |
|---|---|---|---|---|---|
| *Activity Tracking* | 3.17 | 0.27 | 0.000 | 2.62 - 3.72 | 23.91 |
| *Change* | 0.32 | 0.32 | 0.779 | -0.54 - 0.72 | 1.09 |
| *Diary* | 2.67 | 0.77 | 0.001 | 1.15 - 4.19 | 14.53 |
| *Engagement* | 0.55 | 0.55 | 0.041 | 0.04 - 2.22 | 3.11 |
| *Entertainment* | 0.27 | 0.27 | 0.072 | -0.04 - 1.03 | 1.64 |
| *Game Labels* | 0.45 | 0.45 | 0.000 | 2.33 - 4.12 | 25.38 |
| *Player Aspects* | 0.48 | 0.48 | 0.328 | -1.41 - 0.47 | 0.62 |
| *Progress* | 0.39 | 0.39 | 0.075 | -0.07 - 1.48 | 2.03 |
| *Purpose* | 0.67 | 0.67 | 0.186 | -2.21 - 0.43 | 0.40 |
| *Quest* | 0.32 | 0.32 | 0.092 | -1.18 - 0.09 | 0.57 |
| *Quizzes* | 0.64 | 0.64 | 0.000 | 1.94 - 4.45 | 24.44 |
| *Routine* | 1.42 | 0.95 | 0.134 | -0.43 - 3.28 | 4.14 |
| *Statistics* | 1.34 | 0.56 | 0.018 | 0.23 - 2.46 | 3.84 |
| *Story* | 0.96 | 0.43 | 0.025 | 0.12 - 1.80 | 2.62 |
| *Constant* | -2.85 | 0.19 | 0.000 | -3.23 - -2.47 | 0.05 |

OR: Odds Ratio

## IV. DISCUSSION

*A. Context*

The commercial and academic fields show that gamification in health and fitness apps has become common [14], [25]–[27]. A significant amount of research on gamification and health has taken place focusing on physical activity [28]–[32]. Lister et al [14] did a comprehensive review of gamification use in health and fitness apps which showed the need for further examination through large sample studies. However, without a standardized method for detecting gamification, further studies require a manual approach for screening gamification.

Gamification is a motivational design problem that can be solved with design thinking and design processes which requires deep understanding of the users, their motivations and their needs [18]. This understanding will affect how developers and designers might reference elements commonly associated with gamification.

*B. Strengths*

This study represents, to our knowledge, the first automatic method for detecting the presence of gamification in breast cancer and health apps. The importance of this study lies principally in the ability to predict gamification presence using information that is simple to acquire from high-quality secondary sources.

Our method can be used as a screening tool that can easily detect gamification presence in breast cancer apps for further study. Researchers can compare whether the presence of gamification plays a role in the success of an app; researchers can analyze whether correlations exists between the number of downloads, ratings or comments and the presence of gamification. The area under the ROC curve shows that our method's ability to correctly predict gamification presence is above 85%.

Our panel of gamification specialists included members from a variety of disciplines in order to cover different aspects of gamification and make our list of keywords as inclusive as possible. This list is consistent with game elements described by Reeves and Read [33]. These elements included the following: self-representation with avatars; three dimensional environments; narrative context (or story); feedback; reputations, ranks, and levels; marketplaces and economies; competition under rules that are explicit and enforced; teams; parallel communication systems that can be easily configured; and time pressure. The keywords were also in line with gamification components as determined by reviewing the current body of literature and finding common themes and components of gamification used or discussed in the literature for impacting health behavior [11]–[13], [25], [28]. Our two gamification specialists determined that almost one in five of the apps present in our sample contained some kind of gamification elements. Until now, an estimated prevalence of gamification in breast cancer applications was unknown.

We also took the resulting model and developed a working version of our screening tool which can be found online at: https://goo.gl/R7Lv7a. This tool allows users to paste the description of the breast cancer app and get a prediction on the likelihood of it having gamification elements allowing it to be easily repeated and replicated.

*C. Limitations*

The findings of this study should be interpreted in the context of its limitations. There are terms, expressions or concepts that cannot be captured by simple keywords (ie. a text's lusory attitude). Developers may intentionally leave out specific game terms from their descriptions (ie. "winning" at something is not a concept usually used in healthcare). Additionally, this method is based only on app title and description text so gamification elements displayed through user interface or interaction with the app are not accounted. This limitation, however, would also be present for manual operators facing text descriptions only.

While the panel of specialists who created the gamification keyword list covered a broad range of professions, it's possible that the word selection was biased and adding more specialists with different backgrounds will result in a larger or different set.

Logistic regression use as a means to create predictive models has its own set of disadvantages. It relies heavily on having an adequate number of samples for each combination of independent variables. Logistic regression also has an implicit assumption of linearity in terms of the logit function versus the independent variables. Moreover, while the area under the ROC curve of the predicted model was large; the use of an external validation cohort might be needed for evaluating its performance. There are also technical difficulties associated with generating external validation cohorts with similar characteristics, and their usefulness may be limited. There could be variability across groups from other conditions (ie. Type 1 Diabetes, Obesity, etc) and these other conditions may have characteristics that influence text redaction.

V. CONCLUSIONS AND FUTURE RESEARCH

Our study provides researchers an effective screening tool to automatically detect the presence of gamification in breast cancer health applications. Development of future research on this topic can generate more sophisticated tools for detecting gamification presence in health apps. The inclusion of automated analysis of app screenshots will undoubtedly improve its effectiveness and be a great addition to this tool. Including the use of text analytics for keywords extraction would complement our current variable list. The current list of keywords can be expanded by experts in the field and through relevant literature papers text analysis.

ACKNOWLEDGMENT

Guido Giunti gratefully acknowledges the grant number 676201 for the Connected Health Early Stage researcher Support System (CHESS ITN) from the Horizon 2020 Program of the European Commission.

We would like to thank Luis Fernandez-Luque PhD, Jackie Bender MD PhD, Anna Sort RN, Alejandro Lang MBA (PhD), Guido Olomudzski MBA and Dolors Reig LPsy for their help and collaboration.